\documentclass[aps,prl,superscriptaddress,twocolumn,amsmath,amssymb]{revtex4}

\usepackage{color}
\usepackage{graphicx}
\usepackage{bm}
\definecolor{Blue}{rgb}{0.00, 0.00, 1.00}
\definecolor{Red}{rgb}{1.00, 0.00, 0.00}

\begin{document}

\title{Superluminal plasmons with resonant gain in population inverted bilayer graphene}

\author{Tony Low}
\email{tlow@umn.edu}
\affiliation{Department of Elec. \& Comp. Engineering, University of Minnesota, Minneapolis, MN 55455, USA}
\author{Pai-Yen Chen}
\affiliation{Department of Elec. \& Comp. Engineering, Wayne State University, Detroit, Michigan 48202, USA}
\author{D. N. Basov}
\affiliation{Department of Physics, Columbia University, 538 West 120th Street, New York, New York 10027, United States}

\date{\today}
\begin{abstract}

\textcolor{black}{
AB-stacked bilayer graphene with a tunable electronic bandgap in excess of the optical phonon energy presents 
an interesting active medium, and we consider such theoretical possibility in this work. We argue the possibility of a highly resonant optical gain in the vicinity of the asymmetry gap. Associated with this resonant gain are strongly amplified plasmons, plasmons with negative group velocity and superluminal effects, as well as directional leaky modes. 
}

\end{abstract}
\maketitle



\textcolor{black}{
Graphene has emerged as one of the most promising platform for exploring plasmonic effects and light-matter interactions in the mid-infrared spectrum\cite{low2017polaritons,basov2016polaritons,grigorenko2012graphene,koppens2011graphene,low2014graphene,garcia2014graphene}, due to its extreme light confinement, high quality factor\cite{woessner2015highly}, and electrical tunability\cite{chen2012optical,fei2012gate}. An exciting direction is the possibility of imparting gain to graphene plasmon via optically induced population inversion\cite{ryzhii2007negative,rana2008graphene,berini2012surface,stauber2012plasmons,page2015nonequilibrium}. Realization of such plasmonic gain medium would enable tunable active devices such as lossless plasmons\cite{khurgin2012practicality} and plasmonic lasers\cite{bergman2003surface,zheludev2008lasing} in the highly sought after mid-infrared spectrum.}
However, due to the short-lived nature of this inverted state\cite{winzer2012impact,rana2007electron,tomadin2013nonequilibrium,kim2011relaxation,wang2010ultrafast,yan2009time,kang2010lifetimes,george2008ultrafast}, experimental evidence of active plasmon in graphene has so far been elusive\cite{ni2016ultrafast}. 
Upon excitation with an intense short optical pulse in graphene, highly non-equilibrium electrons and holes are generated, and then thermalize rapidly within $10$\,fs into a quasi-equilibrium inverted carrier plasmas\cite{gierz2013snapshots,johannsen2013direct,li2012femtosecond,dawlaty2008measurement}. The lifetime of this inverted state in graphene is limited to around $100\,$fs due to ultrafast carrier recombination processes such as Auger recombination\cite{winzer2012impact,rana2007electron,tomadin2013nonequilibrium}, optical phonons\cite{kim2011relaxation,wang2010ultrafast,yan2009time,kang2010lifetimes} and plasmons emissions\cite{george2008ultrafast}. As a result of this ultrashort-lived transient state, a consequence of the absence of an electronic gap, the observation of optical gain (negative conductivity)\cite{ryzhii2007negative} associated with the inverted state has been elusive in experiments.


AB-stacked bilayer graphene\cite{mccann2006asymmetry,ohta2006controlling}, with a tunable electronic bandgap up to $300\,$meV\cite{zhang2009direct}, offers an interesting active medium for several reasons. First, an electronic gap in excess of the optical phonon energy ($\sim 0.2\,$eV in graphene) would suppress non-radiative carrier recombinations due to energy conservation\cite{basko2008theory}. Second, since energy and momentum must be simultaneously conserved,  Auger events are in general less likely to occur with increasing bandgap\cite{delaney2009auger}. Lastly, pertaining to non-radiative loss via plasmons, we note that spontaneous emission of plasmons is not necessarily a nuisance, as stimulated emission also implies plasmon amplification\cite{rana2008graphene,berini2012surface,stauber2012plasmons,page2015nonequilibrium}, the topic of interest in this work. 

\begin{figure*}[t]
\centering
\scalebox{0.7}[0.7]{\includegraphics*[viewport=10 190 600 750]{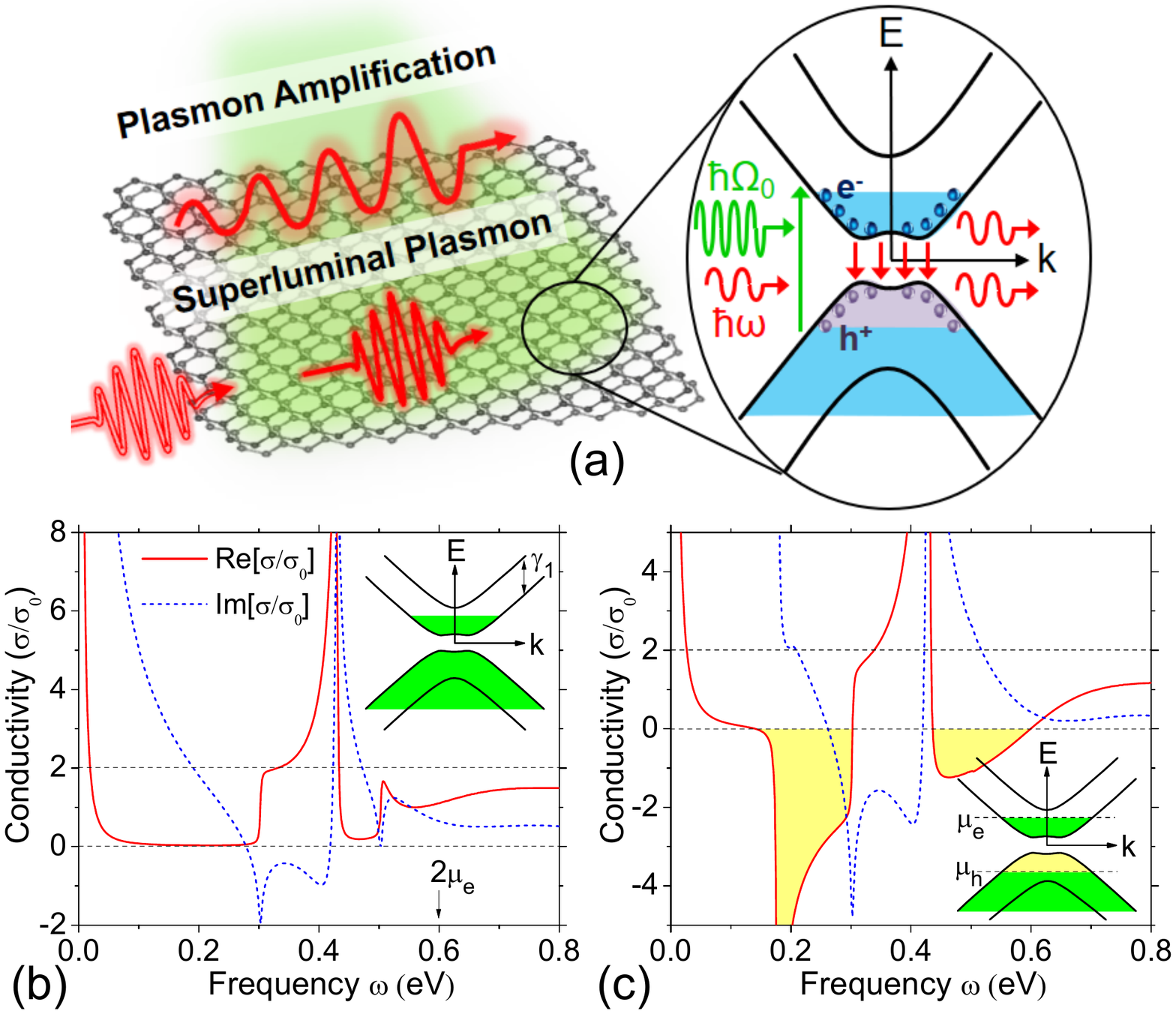}}
\caption{\textbf{Optical conductivity.} \textbf{(a)} Schematic illustrating the singularity in joint optical density-of-states in gapped bilayer graphene under population inversion. This produces a resonant optical gain, leading to amplified plasmons with superluminal effects. 
\textbf{(b)} Complex optical conductivity of AB-stacked bilayer graphene with an asymmetry gap, $\Delta=0.2\,$eV. Carriers assumed to be in equilibrium with $\mu_e=\mu_h=0.3\,eV$, referenced with respect to the Dirac point.
Calculation assumed $T=300\,$K and  $\eta\approx 10\,$meV.
\textbf{(c)} Similar to (b), but for the inverted state,  $\mu_e=-\mu_h=0.3\,eV$. Resonant optical gain in vicinity of $\Delta$ is observed, due to singularity in the joint density-of-states from the Mexican hat bandstructure (see inset).
}
\label{fig1}
\end{figure*}

\textcolor{black}{Although the regime of sufficiently long-lived optical gain for bilayer graphene has not been experimentally realized yet, we consider this a theoretical possibility for the above-mentioned reasons. }
In this work, we will examine the attributes of gapped bilayer graphene under population inversion as a plasmonic gain media. We predict the existence of a resonant optical gain at frequencies around the energy gap due to a singularity in its joint optical density-of-states as illustrated in Fig.\,\ref{fig1}a. Associated with the resonant optical gain are highly amplified transverse magnetic plasmonic mode and an accompanying spectral window with negative group velocity exhibiting superluminal effects. Besides the predictions of these novel near-field effects, we also found directional active leaky modes into far-field. Since these effects rely on the opening of an electronic gap, they are therefore highly electrically tunable. We also discuss relevant experimental considerations for their possible realization.

\emph{Model Hamiltonian---} 
Bilayer graphene arranged in the Bernal AB stacking order is considered,
with basis atoms $A_1$, $B_1$ and  $A_2$, $B_2$ in the top and bottom layers respectively.
The intralayer coupling is $\gamma_0\approx 3\,$eV  and the interlayer coupling between
$A_2$ and $B_1$ is  $\gamma_1\approx 0.39\,$eV,
an average of values reported in infrared  and photoemission measurements\cite{kuzmenko2009infrared,wang2008gate,li2009band,ohta2006controlling,zhou2008origin}.
We work within the $4\times 4$ atomic $p_z$ orbitals basis, i.e.
$a^{\dagger}_{1\bold{k}},b^{\dagger}_{1\bold{k}},a^{\dagger}_{2\bold{k}},b^{\dagger}_{2\bold{k}}$,
where $a^{\dagger}_{i}$ and $b^{\dagger}_{i}$ are creation operators for the $i^{th}$
layer on the $A/B$ sublattices.
Within this basis, the Hamiltonian near the $\bold{K}$ point can be written as:
${\cal H}_{k}=v_f \pi_{+}I\otimes\sigma_{-}+v_f \pi_{-}I\otimes\sigma_{+}+\tfrac{\Delta}{2}\sigma_z\otimes I+\gamma_1/2[\sigma_x\otimes\sigma_x+\sigma_y\otimes\sigma_y]$,
where $\sigma_i$ and $I$ are the Pauli and identity matrices respectively.
We defined $\sigma_{\pm}\equiv\tfrac{1}{2}(\sigma_x \pm i \sigma_y)$
and $\pi_{\pm}\equiv\hbar(k_x\pm ik_y)$. Here, $v_f$ is the Fermi velocity and $\Delta$ is the asymmetry gap due to electrostatic potential
difference between the two layers. Expressions for non-interacting ground state
electronic bands $\xi_{n}(\bold{k})$ ($n=1-4$, see insets of Fig.\,\ref{fig1}) and wavefunctions $\Phi_n(\bold{k})$
are obtained by diagonalizing ${\cal H}_{k}$.


\emph{Optical conductivity---} 
The optical conductivity at equilibrium can be obtained from the
Kubo formula as function of frequency and momentum, which reads,
\begin{eqnarray}
\sigma_{\alpha\alpha}(\textbf{q},\omega)=-i\frac{g\hbar e^2}{(2\pi)^2}\sum_{nn'} \int d\textbf{k}\: \: \: \: \: \: \: \: \: \: \: \:\\
\nonumber
\times \frac{f(E_{n\textbf{k}}-\mu)-f(E_{n'\textbf{k}'}-\mu)}{E_{n\textbf{k}}-E_{n'\textbf{k}'}}
\frac{|\left\langle \Phi_{n\textbf{k}}\right|\hat{v}_{\alpha}\left|\Phi_{n'\textbf{k}'}\right\rangle|^2 }{E_{n\textbf{k}}-E_{n'\textbf{k}'}+\hbar\omega+i\eta}
\label{kubofor}
\end{eqnarray}
where $e$ is the electronic charge, $\hat{v}_{\alpha}$ is the velocity operator defined as $\hbar^{-1}\partial_{k\alpha}{\cal H}$, $g=4$ accounts for the spin and valley degeneracies and $\eta\approx 10\,$meV accounts for the finite damping. $f(...)$ is the Fermi-Dirac distribution function, $\mu$ is the Fermi level, and temperature is taken to be $300\,$K in all calculations. The indices $n$ denotes the band indices. $E_{n\textbf{k}}$ and $\Phi_{n\textbf{k}}$ are the eigen-energies and eigen-functions of ${\cal H}$.  For the inverted state, we replace $f(...)$ with a quasi-equilibrium distribution given by, 
\begin{eqnarray}
n_F(E_{n\textbf{k}})= \theta(E_{n\textbf{k}})f(E_{n\textbf{k}}-\mu_e)+ \theta(-E_{n\textbf{k}})f(E_{n\textbf{k}}-\mu_h)
\end{eqnarray}
Here, we assume that the quasi-Fermi levels are such that $\mu_h=-\mu_e$, and that the electron and hole baths can be described by a common temperature. Optical pumping, where electrons and holes are generated in pairs, of a charge neutral system with particle-hole symmetry would fit such scenario.

Fig.\,\ref{fig1}b shows the complex optical conductivity of doped bilayer graphene with an asymmetry gap of $\Delta=0.2\,$eV, under equilibrium. Although the equilibrium case has been discussed in the literature\cite{nicol2008optical,zhang2008determination,li2009band,kuzmenko2009infrared,benfatto2008robustness}, it is instructive to review the main features. The main departure from that of monolayer graphene spectra, which is well-known\cite{basov2014colloquium}, is the appearance of a resonant conductivity peak at $\omega\sim\gamma_1$. This can be traced to the two conduction bands separated by $\gamma_1$, which in the presence of a finite $\Delta$ distorts the otherwise perfect nesting. The broadening of the peak into a broad shoulder of width  $\sim\tfrac{\Delta}{2}$  is related to the gap opening. A weaker feature at $\omega\sim 0.5\,$eV can be observed, and is due to optical transitions between the Mexican-hat valence band to the excited conduction band.

\emph{Resonant gain---}
Fig.\,\ref{fig1}c shows the complex optical conductivity for the non-equilibrium population inverted state. In the case of graphene, the population inversion entails that the optical conductivity changes sign for $\omega\lesssim\mu_e-\mu_h$\cite{ryzhii2007negative,li2012femtosecond}. The opening of an asymmetry gap $\Delta$  implies no optical transitions between the inverted populations at $\omega\lesssim\Delta$. Hence, this impose a lower cutoff, and the negative conductivity regime should be given by  $\Delta\lesssim\omega\lesssim\mu_e-\mu_h$ instead. Due to the singularity in the joint density-of-states at $\sim \Delta$, we obtain resonant gain. This expectation is largely reflected in Fig.\,\ref{fig1}c, except that the resonant absorption at $\gamma_1$ due to the nested bands still survives. The latter does not translate to gain as its 
optical transitions are not between bands of inverted population. In fact, if one compares the shape of the peak at $\gamma_1$ to the broad negative conductivity background, it suggests a roughly two-fold increase in its spectral weight relative to the equilibrium case in Fig.\,\ref{fig1}b. This enhancement in the $\gamma_1$ peak in the population inverted case is due to the additional hole contributions between the nested valence bands. Since, the optical processes are electron-hole symmetric, we simply get a two-fold increase in spectral weight.

\emph{Plasmon dispersion---}
The plasmonic response of bilayer graphene can be obtained from its
dielectric function given by,
$\epsilon_T^{rpa}(q,\omega)=\kappa-v_{c}\Pi(q,\omega)$,
at arbitrary wave-vector $q$ and frequency $\omega$, and
$v_c=e^2/2q\epsilon_0$ is the $2D$ Coulomb interaction.
\textcolor{black}{
$\Pi(q,\omega)$ is the non-interacting part (i.e. the pair bubble diagram) of
the charge-charge correlation function given by\cite{giuliani2005quantum,wunsch2006dynamical,hwang2007dielectric}, and is
related to the conductivity via $\sigma=i\tfrac{e^2\omega}{q^2}\Pi$.}
The plasmon dispersion $\omega_{pl}(q)$ is given by the zeros of the dynamical dielectric function. The loss function, defined as $L(q,\omega)=-\mbox{Im}[1/\epsilon(q,\omega)]$, provides a direct measure of the spectral weight of elementary excitations. In the limit of zero loss, $L(q,\omega)\sim \delta(\omega-\omega_{pl}(q))$. Alternatively, the plasmon dispersion can also be obtained by solving the source-free Maxwell equations for the transverse magnetic (TM) surface wave,
\begin{eqnarray}
q\equiv \beta + i\alpha= \frac{\omega}{c} \sqrt{1- \left(\frac{2}{\sigma(\omega)\eta_0}\right)^2}
\label{tm}
\end{eqnarray}
where $\eta_0=\sqrt{\mu_0/\epsilon_0}$ is the free space impedance. Here, the choice of signs for $\alpha$ and $\beta$ are related to the active/passive conductivity, ensuring that there is no violation of passivity and causality\cite{collin1969antenna}.

\begin{figure*}[!htbp]
\centering
\scalebox{0.72}[0.72]{\includegraphics*[viewport=110 20 465 790]{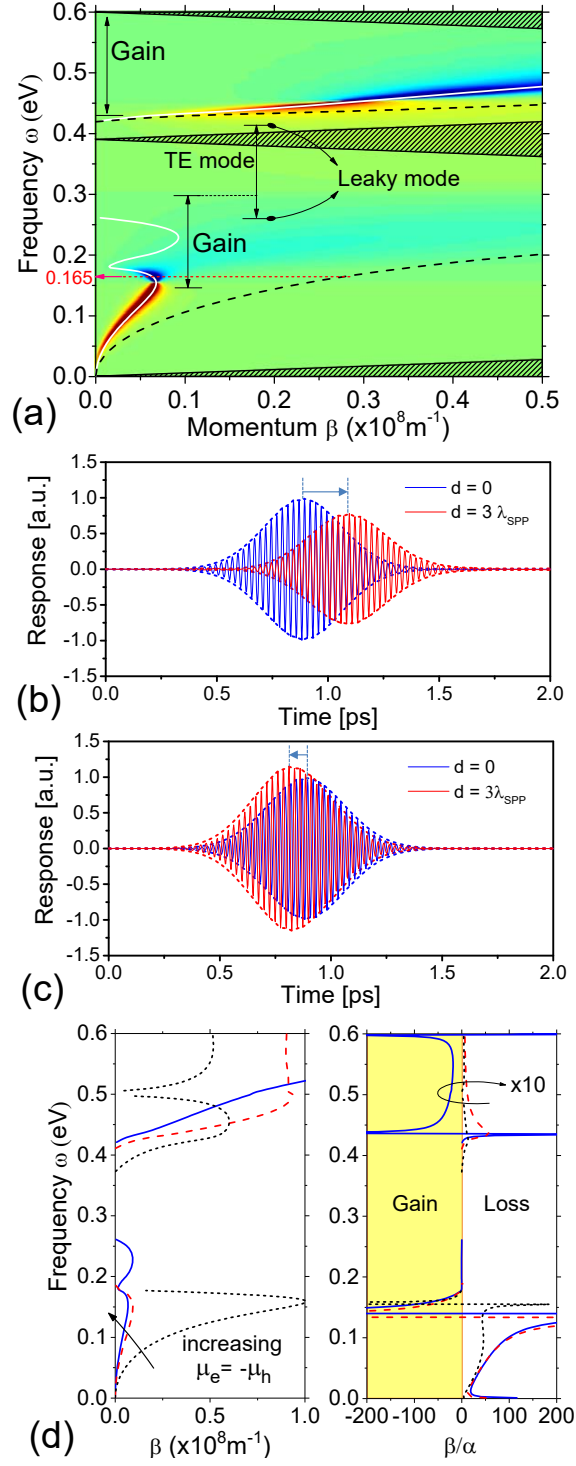}}
\caption{\textbf{Plasmon.} \textbf{(a)} Loss function for bilayer graphene with an asymmetry gap, $\Delta=0.2\,$eV, and for the inverted state,  $\mu_e=-\mu_h=0.3\,eV$.
Calculation assumed $T=300\,$K and  $\eta\approx 10\,$meV.
For the color scale, red (blue) denotes loss (gain).
Plasmon dispersions for the equilibrium (black dashed) and inverted (white solid) state are calculated using Eq.\,\ref{tm} using the optical conductivity displayed in Fig.\,\ref{fig1}.
Shaded regions are electron-hole continua ($\mbox{Im}[\epsilon(q,\omega)]\neq 0$) where plasmons mode would be Landau damped.
\textbf{(b)} Transient responses of a short plasmon pulse with central frequency at $0.165\,$eV in bilayer graphene under equilibrium, initiated at position $x=0$ and then recorded at several plasmon wavelengths. Results indicate a positive group delay. 
\textbf{(c)} Similar to (b), but for the case with population inversion, revealing a negative group delay due to the negative group velocity. 
\textbf{(d)}
Complex dispersion of plasmon, $\beta(\omega)$ and $\alpha(\omega)$, for the inverted state for varying dopings of $\mu_e=-\mu_h=0.1$ (black dotted), $\,0.2$ (red dashed), $\,0.3\,eV$ (black solid). The quality factor $\beta/\alpha$ is $\times 10$ for better readability across the spectral range $\omega>0.4\,$eV.
}
\label{fig2}
\end{figure*}

We proceed to examine the plasmons in photo-inverted bilayer graphene. Using the optical conductivity obtained in Fig.\,\ref{fig1} for the equilibrium and inverted state, we compute their plasmon dispersions as shown in Fig.\,\ref{fig2}a as black dashed and white solid lines, respectively. Besides the conventional $\omega_{pl}\sim\sqrt{\beta}$ plasmons\cite{fei2015tunneling,yan2014tunable}, it is known theoretically that bilayer graphene also accommodates an over-damped acoustic mode ($\omega_{pl}\sim \beta$)\cite{sensarma2010dynamic,gamayun2011dynamical} 
, and a high energy $\gamma$ plasmon mode ($\omega_{pl}-\gamma\sim \beta$)\cite{low2014novel}. The former and latter are clearly depicted in Fig.\,\ref{fig2}a. Besides the obvious larger stiffness in the plasmon modes for the inverted state due to the increased Drude weight, we also observe strong departure from the $\sqrt{\beta}$ dispersion. In spectral regions of optical gain, $\mbox{Re}(\sigma)$$<$$0$ as indicated in Fig.\,\ref{fig2}a, the $\sqrt{\beta}$ mode can acquire anomalous backwards bending dispersion. Plasmon gain can also be revealed through the negative spectral weights in the loss function, similarly for the $\gamma$ plasmon mode.
\textcolor{black}{We note that there has been several theoretical proposals for exploiting joint density-of-states singularities for plasmon emission and gain\cite{vitlina2003amplification,enaldiev2017plasmon,svintsov2016plasmons}, albeit mostly in resonant tunneling structures.}

\emph{Superluminal effects}
This plasmonic spectral region of negative group velocity with gain is highly nontrivial. For waves propagating in passive dispersive media, the group velocity can become negative in presence of strong resonant absorption\cite{brillouin2013wave}, however, the accompanying losses has impeded their direct experimental observation. On the other hand, if the optical medium has resonant gain (i.e. inverted populations) instead of absorption, negative group velocity has been experimentally observed\cite{wang2000gain,bigelow2003superluminal}.
Gapped bilayer graphene, with its resonant optical gain, can allow for similar exploration of superluminal effects associated with negative group velocities in the rarely explored mid-infrared regime.

First, consider a pulse of TM surface wave with central frequency $0.165\,$eV propagating along the bilayer graphene under equilibrium, i.e. Fig.\,\ref{fig1}b. The pulse is initiated at position $x=0$, and then recorded at $x=3\lambda_{spp}$, where $\lambda_{spp}=2\pi/\beta$ is the wavelength of the surface plasmon wave. Per the dispersion relation in Fig.\,\ref{fig2}a, the group velocity of passive bilayer graphene is positive, and slower than the speed of light. In this case, it can be seen from Fig.\,\ref{fig2}b that the output signal lags far behind the input, with a propagation time of $0.22\,$ps for the peak intensity. For the same distance, the light propagation in free-space would have been $2\,$fs instead. Since $\alpha>0$, the output signal amplitude is reduced as it passes through the lossy medium. 

Fig.\,\ref{fig2}c shows the similar time domain response for the population inverted bilayer graphene. In this case, the central carrier frequency $0.165\,$eV now resides in the superluminal band as depicted in Fig.\,\ref{fig2}a, with the negative slope dispersion (i.e. $d\beta/d\omega<0$). It can be observed that the peak envelop of the output leads that of the input signal, hence, the transit time of the signal is effectively negative (i.e. $\sim -0.047\,$ps), or superluminal. At the same time, the pulse intensity is also gradually amplified due to the optical gain in population inverted bilayer graphene. We should emphasize here that the observed superluminal pulse propagation does not violate causality\cite{chiao1993superluminal,mitchell1998causality} and can be explained by a direct consequence of the interference between its different components in an active, anomalously dispersive medium. Alternatively, one can also see that although the peak envelop of the signal is superluminal, the front end of the signal does not\cite{gauthier2007fast}.


\emph{Effect of photo-doping---}
Fig.\,\ref{fig2}d explores the effect of photo-dopings, i.e. $\mu_e=-\mu_h=0.1,\,0.2,\,0.3\,eV$, on the complex plasmon dispersion, $k\equiv \beta+i\alpha$. Increased stiffness in $\beta$ is simply due to the increased Drude weights. Recall that the existence of TM modes requires that $\mbox{Im}(\sigma)$$>$$0$, otherwise, we have the weakly confined transverse electric (TE) modes. Hence, region of spectral gap for the TM modes is observed, whose gap increases with doping. Abrupt switching of the TM modes from damped ($\alpha>0$) to amplification ($\alpha<0$) is observed when gain is introduced, i.e. $\mbox{Re}(\sigma)<0$. Spectral bandwidth of amplification increases with doping, and at sufficiently large doping, this amplification can also be imparted to the $\gamma$ plasmon mode\cite{low2014novel} at energy $\sim\gamma_1$.  
The quality factor, conventionally defined as $\beta/\alpha$, where $\left|\beta/\alpha\right|/2\pi$ gives the number of cycles the plasmon can propagates before its amplitude decays by $1/e$. In the presence of gain, $\beta/\alpha$ instead gives the number of cycles before the amplitude gets amplified by $e$. Fig.\,\ref{fig2}d shows that the plasmon mode can be strongly amplified in fraction of a propagation cycle, particularly within spectral region of resonant gain. In reality, the amplitude cannot grow indefinitely, as its energy cannot exceeds that being input into the system e.g. via optical or electrical pumping. Nonlinear effects will begin to dominate.

\emph{Leaky modes---}
As highlighted in Fig.\,\ref{fig2}a, in the vicinity of the transition between the TM and TE modes (i.e. $\mbox{Im}(\sigma)\approx 0$), $\beta$ (TM mode) can be less than the free-space wavevector $k_0$ in a narrow frequency range. Such a ``fast" wave can easily out-couple into free-space, i.e. so called leaky mode. From Fig.\,\ref{fig2}a, we note that the leaky mode at energy $\sim 0.26\,$eV is active as it overlaps with the gain region, while another leaky mode at higher energy is passive. These observations can also be verified from complex wave-vector in Fig.\,\ref{fig2}d. Indeed, 
for the leaky mode at $\sim 0.26\,$eV, $\left| \alpha \right|$ is large, and at the same time $\beta<k_0$.

\begin{figure*}[t]
\centering
\scalebox{0.7}[0.7]{\includegraphics*[viewport=140 200 465 760]{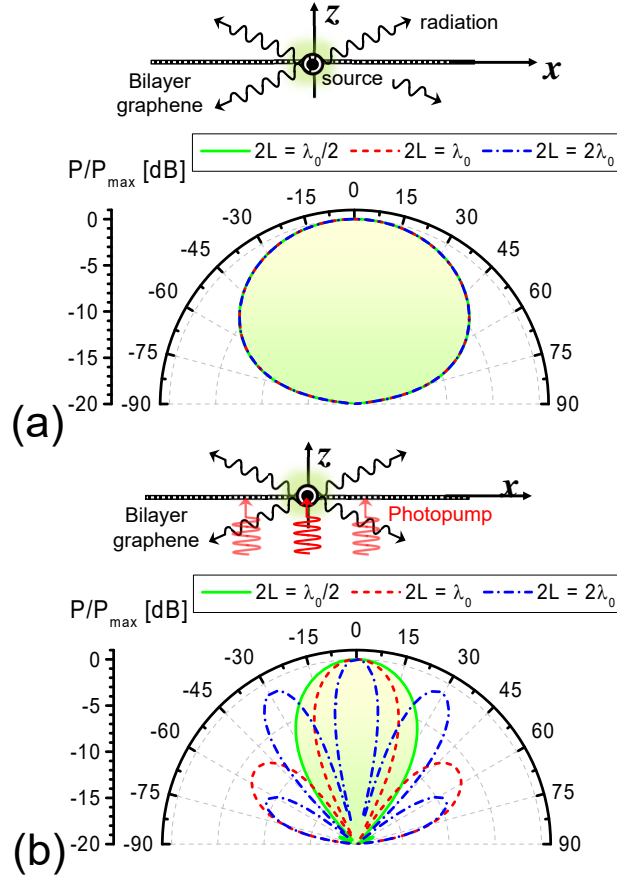}}
\caption{\textbf{Leaky modes.}  \textbf{(a)} Radiation patterns for the graphene bilayer excited by a magnetic line source at photon energy of $0.42\,eV$, varying the length of graphene sheet. \textbf{(b)}  Similar to (a), but for the pumped graphene bilayers at photon energy of $0.26\,eV$ (active leaky-mode regime). }
\label{fig3}
\end{figure*}

Such leaky mode, when excited by a source, produces electromagnetic radiation into free space. This implies that energy optically pumped into the bilayer graphene might escape into free-space via these modes. Consider bilayer graphene with finite length of $2L$, excited by an infinite magnetic line source placed at its center (see inset of Fig.\,\ref{fig3}). 
The far-field radiation pattern with respect to the out-of-plane axis can be derived as\cite{jackson1988leaky},
\begin{eqnarray}
P(\theta)\propto \left| \mbox{cos}(\theta)\frac{\Omega}{(\alpha\mp i\beta)^2+k_0^2\mbox{sin}^2\theta}          \right|^2
\label{leaky}
\end{eqnarray}
and
\begin{eqnarray}
\nonumber
\Omega\equiv \alpha\mp i\beta- \mbox{exp}[-L(\alpha\mp i\beta)]\times\\
\left[    (\alpha\mp i\beta)\mbox{cos}(k_0 L \mbox{sin}\theta)  -k_0 \mbox{sin}\theta \mbox{sin}(k_0 L \mbox{sin}\theta)     \right]
\label{leaky2}
\end{eqnarray}
where $\mp$ denotes the sign of $-\mbox{Re}(\sigma)$. Using the complex dispersion obtained in Fig.\,\ref{fig2}, we analyze  the leaky modes under equilibrium and inverted population.

Fig.\,\ref{fig3}a plots the far-field radiation pattern for bilayer graphene under equilibrium at photon energy of $0.42\,$eV, corresponding to a passive leaky mode highlighted in Fig.\,\ref{fig2}a. Since $\alpha$ is quite large, the radiation pattern reveals a 
large beam width, with maximum in the broadside ($\theta=0$) and zero radiation in the end-fire direction ($\theta=\pm\pi$). These observations can also be shown from Eq.\,\ref{leaky}, which in the $\alpha\gg\beta,k_0$ limit, $P\sim \mbox{cos}^2\theta/\alpha^2$.
For  $L$ sufficiently larger than the plasmon attenuation length (usually much shorter than free-space wavelength), the radiation pattern becomes $L$-independent. 

Radiation pattern for the inverted bilayer graphene, where plasmons are being amplified along the propagation direction, is a strong function of $L$ instead. These results are shown in Fig.\,\ref{fig3}b. In the active leaky-mode region, 
the amplified plasmon mode results in different resonant radiative states as function of $L$, analogous to elongated filament currents of different lengths. As $L$ increases, the maximum radiation at broadside, and the number of radiation lobes also increases, while the beam width of the main beam (broadside) decreases. The active leaky modes can produce more directive main beam and richer radiative pattern, when compared to its passive counterpart of same $L$. Such tunable leaky waves may find applications in infrared nanoantenna and emitters.


\emph{Concluding remarks---} We discussed several interesting near- and far-field midinfrared responses of gapped bilayer graphene under population inversion. In particular, we identified a resonant optical gain in vicinity of the asymmetry gap, and 
associated with this are strongly amplified plasmons with nontrivial dispersion, such as negative group velocity. This will be especially interesting in terms of nano-imaging\cite{basov2016polaritons,low2017polaritons,fei2012gate,chen2012optical} observables, since both the sign and magnitude of plasmon phase and group velocities are readily measurable\cite{yoxall2015direct}.
For waves propagating through negative group velocity medium, scenarios of superluminal can also occurs. For example, the peak of a temporally long Gaussian pulse can appear at the rear side before the peak even enters the medium, which is not at conflict with relativity or causality\cite{chiao1993superluminal}.
\textcolor{black}{
Our predictions in this work largely rely on the huge joint density-of-states in gapped bilayer graphene. Experimentally, signatures of such singularities have already been observed via infrared measurement\cite{kuzmenko2009infrared}, quantum capacitance\cite{young2012electronic}, and scanning tunneling microscope\cite{brar2007scanning}. However, the quality of the van Hove singularity is largely smeared by disorder such as charge impurities. The recent advent of ultra high quality bilayer graphene encapsulated with hexagonal boron nitride would be a promising platform\cite{ju2017tunable}.}






\emph{Acknowledgements---} TL acknowledges support from NSF through the University of Minnesota MRSEC under Award Number DMR-1420013. DNB is the Moore Foundation Investigator, EPiQS Initiative GBMF4533. TL and DNB acknowledge support from the National Science Foundation under grant number NSF/EFRI-1741660. PYC acknowledges support from the National Science Foundation under grant number NSF/ ECCS-1711409 and KAUST Competitive Research Grant Program under grant number CRG-2953.

\newpage

\newpage

\newpage

\end{document}